\documentclass[letter,12pt]{article}
\usepackage[letterpaper,margin=1in]{geometry}
\usepackage{authblk}
\usepackage{amsmath}
\usepackage{amsfonts}
\usepackage{rotating}
\usepackage{booktabs}
\usepackage[labelfont=bf]{caption}
\usepackage{setspace} 
\usepackage[english]{babel}
\usepackage{graphicx}
\usepackage[document]{ragged2e}
\usepackage{times}
\usepackage{algpseudocode}
\usepackage{amssymb}
\usepackage{bbm}
\usepackage{xcolor}
\usepackage{multirow}

\usepackage{lineno}

\usepackage{array}
\newcolumntype{P}[1]{>{\centering\arraybackslash}p{#1}}

\usepackage{amsmath}
\usepackage{hyperref}
\usepackage[style=nature]{biblatex}
\renewcommand\Affilfont{\fontsize{11}{12}\itshape}
\makeatletter
\renewcommand\AB@affilsepx{; \protect\Affilfont}
\makeatother

\title{\Large \bf Bias amplification in experimental social networks \\ is reduced by resampling}

\author[1,*,\textdagger]{Mathew D. Hardy}
\author[2,\textdagger]{Bill D. Thompson} 
\author[3]{P.M. Krafft}
\author[1,4]{Thomas L. Griffiths}

\affil[1]{Princeton University, Department of Psychology, Princeton, NJ, USA} 
\affil[2]{Department of Psychology, University of California, Berkeley, Berkeley, CA, USA}
\affil[3]{Creative Computing Institute, University of the Arts London, London, UK}
\affil[4]{Princeton University, Department of Computer Science, Princeton, NJ, USA}
\affil[*]{Corresponding author: Mathew Hardy, \texttt{mdhardy@princeton.edu}; \textsuperscript{\textdagger}These authors contributed equally}

\date{}

\begin{document}
\begin{refsection}
\maketitle
\justifying
{\bf 
Large-scale social networks are thought to contribute to polarization by amplifying people's biases. However, the complexity of these technologies makes it difficult to identify the mechanisms responsible and to evaluate mitigation strategies. Here we show under controlled laboratory conditions that information transmission through social networks amplifies motivational biases on a simple perceptual decision-making task. Participants in a large behavioral experiment showed increased rates of biased decision-making when part of a social network relative to asocial participants, across 40 independently evolving populations. Drawing on techniques from machine learning and Bayesian statistics, we identify a simple adjustment to content-selection algorithms that is predicted to mitigate bias amplification. This algorithm generates a sample of perspectives from within an individual's network that is more representative of the population as a whole. In a second large experiment, this strategy reduced bias amplification while maintaining the benefits of information sharing.
}

\newpage

Large-scale social media platforms are transforming how people communicate and consume information online \supercite{lerman2010information,bakshy2012role,lerman2007social}, shaping media narratives \supercite{hermida2016social}, political discourse \supercite{gainous2013tweeting}, and popular culture \supercite{burns2009celeb}. However, the complex networks created by social media platforms can have unexpected outcomes. For example, social networks often lead to ``echo-chambers'' of like-minded individuals \supercite{bakshy2015exposure, conover2011political,pariser2011filter, levy2021social, cinelli2021echo, shin2017partisan}, raising concerns that social interaction on these platforms can increase polarization and amplify bias \supercite{settle2018frenemies}, facilitating the spread of inaccurate \supercite{allcott2019trends,allcott2017social,tucker2018social}, extreme \supercite{yarchi2021political}, and emotionally charged \supercite{brady2021social} views. 

Understanding platform effects and developing potential solutions is a global priority, leading to calls for better governance of social media platforms \supercite{balkin2020regulate,cusumano2021social}. An important part of these potential solutions is the development of content-selection algorithms that demonstrably reduce harmful platform effects \supercite{lazer2018science}. However, the complexity of large-scale social networks makes it hard to identify specific mechanisms that cause unwanted platform effects and to evaluate the effectiveness of mitigation strategies. These challenges have hindered development of practical frameworks for designing less harmful algorithms \supercite{bail2022social}.

\begin{figure}[t!]
    \centering
    \includegraphics[width=\textwidth]{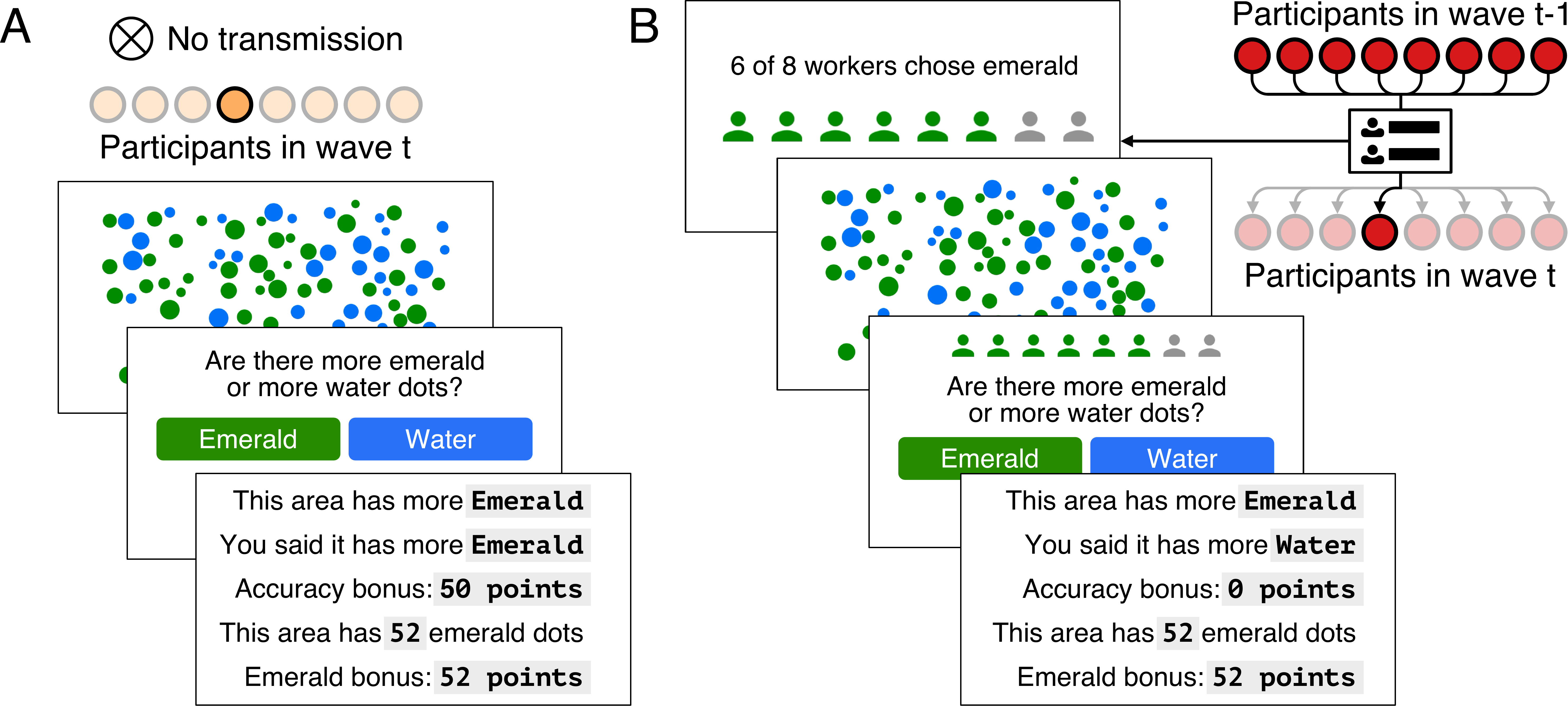}
    \caption{
    \textbf{Perceptual decision-making task}. Our experimental paradigm consisted of arranging participants into an ordered set of groups, called ``waves''. At each wave $t$ participants in social conditions observed judgments made by the participants in wave $t-1$. Participants in asocial conditions did not observe any social information. (A) Asocial/Motivated condition. Each colored circle at the top of the image represents a participant. Stimuli consisted of 100 randomly positioned and sized blue and green dots displayed for one second. After viewing a stimulus, participants indicated whether they thought the stimulus had more green or more blue dots. Participants received feedback after each judgment on practice trials, and at the end of the experiment on test trials. All participants received a bonus on every trial if their judgment was correct. Participants in motivated conditions (shown here) received an additional bonus on every trial for every dot of their motivated color (green in both plots) regardless of whether their judgment was correct. (B) Social/Motivated condition. Each colored circle at the top of the image represents a participant. Solid line arrows indicate flow of information. The perceptual task was the same for participants in asocial and social conditions. However, before and after viewing each stimulus, participants in social conditions observed the aggregate judgments made by participants in the previous wave on the same stimulus. Social information from one wave to the next was presented as the number of participants making the majority judgment (ties were broken randomly).}
    \label{fig:task}%
\end{figure}

One way to address these challenges is to test the effects of different content-selection algorithms in controlled laboratory settings. In this paper, we use an experimental paradigm to study how information sharing affects bias in judgment and decision-making. This new experimental paradigm allowed us to evaluate a mathematical theory of bias amplification and test a mitigation strategy based on this theory. Our approach builds on prior research that has demonstrated benefits \supercite{mason2012collaborative, becker2017network, jayles2017social, rendell2010copy, smaldino2013human} and risks \supercite{moussaid2015amplification, luo2022credibility} of information sharing and that has identified cognitive biases \supercite{kahneman1982judgment} that can be reliably produced in experimental settings. In particular, we examined how social networks can amplify motivated perception, a bias in judgment and decision-making in which people perceive events in a way that supports their desires \supercite{hastorf1954they, dunning2013wishful, leong2019neurocomputational, bruner1947value, balcetis2006see}.

Our experiments used a simple perceptual task: Participants were briefly shown displays of 100 randomly positioned and sized blue and green dots, and were asked to judge whether the stimulus contained more green or more blue dots (see Figure~\ref{fig:task}). Participants received a monetary reward for every correct answer. However, certain participants were offered an additional monetary reward for every green or blue dot in each stimulus (``motivated color'' was randomized across participants, see Materials and Methods, Experiment 1). Because it was earned regardless of a participant's judgment, this reward would not influence the judgments of a rational observer. However, based on previous results \supercite{dunning2013wishful, leong2019neurocomputational} we expected that it would induce a motivated perception bias leading people to overestimate the number of motivated-color dots.

To better understand how social networks impact decision-making in this controlled setting, we designed custom software to construct evolving social networks by recruiting thousands of participants in sequences of discrete groups, or recruitment \emph{waves}. Participants in each wave observed the judgments made by participants in the previous wave before making their own judgments. This created a network of social influence that captured the basic structure of social media: People formed opinions that were informed by the shared opinions of others. Using this controlled experimental approach, we explored the impact of social networks on biases in decision-making and how unwanted impacts could be mitigated.

\section*{Model}

\begin{figure}[t!]
\centering
\includegraphics[width=\textwidth]{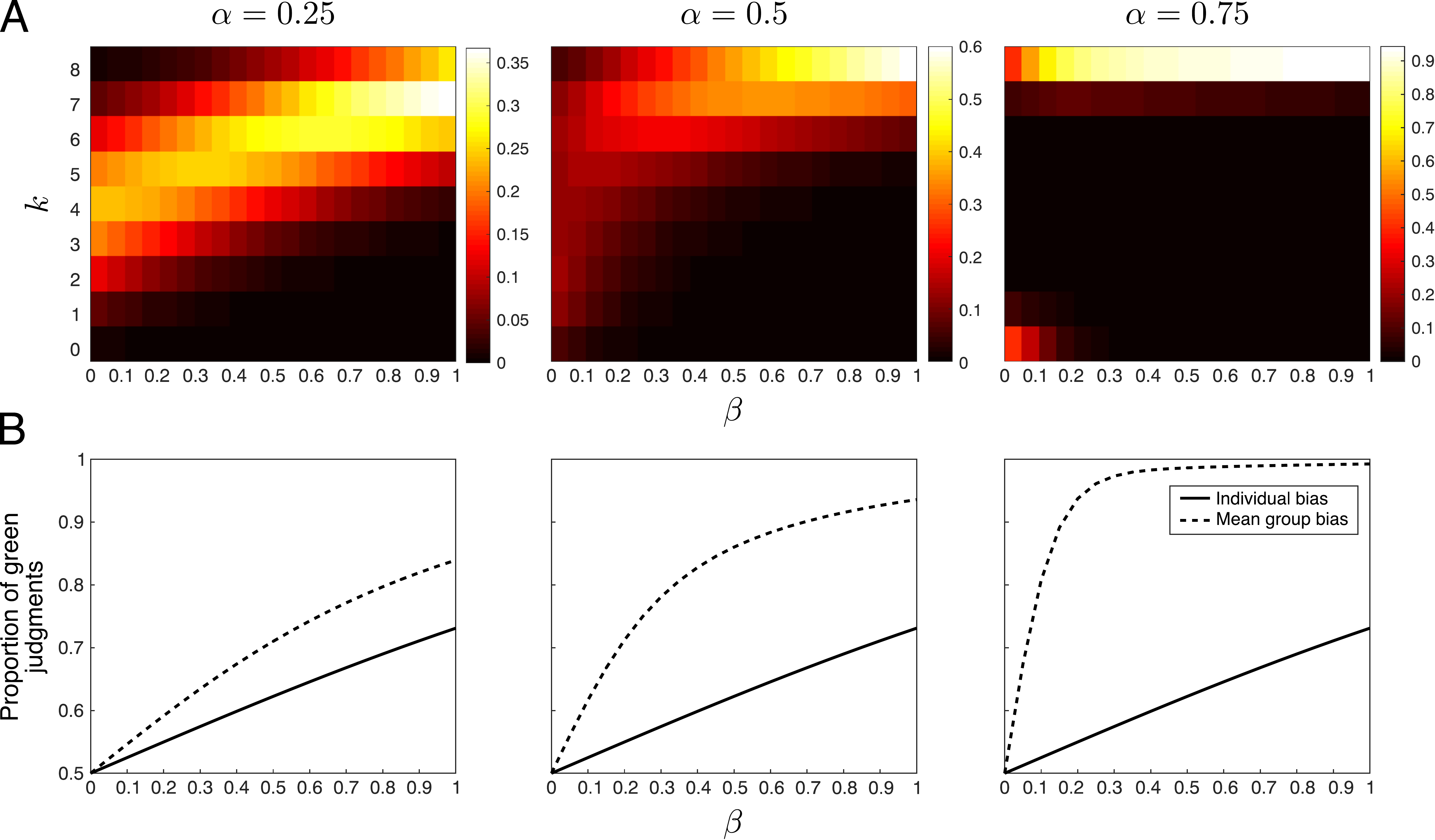}
\caption{\textbf{Stationary distributions for social networks.} (A) Stationary distributions on the number of people endorsing green as a function of the bias $\beta$ towards green, for different levels of sensitivity to social information $\alpha$. The bias translates into a stationary distribution strongly skewed towards green, with increasing effect as $\alpha$ increases. (B) Average proportion of green judgments under this stationary distribution, compared against the bias of a single individual. The social network amplifies individual biases. Because bias $\beta$ and the information from the stimulus $\gamma_d$ have the same impact on judgments, this model predicts that the effects of both will be exaggerated by participating in a social network: people will become more biased, but also more accurate. Note that this analysis could be equivalently stated in terms of blue bias and blue judgments.}
\label{fig:statdist}
\end{figure}

In order to generate quantitative predictions about the consequences of information sharing in this context, we performed a formal analysis of the hypothesis that social networks can amplify people's biases. First, we defined a simple Bayesian model of the decision-making process. Participants were asked to evaluate two possible hypotheses $h$---that there are more green dots ($h=g$) or more blue dots ($h=b$). This judgment was based on the observed perceptual data $d$. It was also potentially based on the social signal $s$, in which $k$ out of $n$ other people judged that there were more green dots. The posterior probability of choosing green $p(h=g|d,s)$ can be computed by applying Bayes' rule. In the Materials and Methods we show that this probability can be written as a logistic function,
\begin{equation}
    p(h=g|d,s) = \frac{1}{1+\exp \{ -\alpha(k-n/2)  -\gamma_d - \beta \}}, \label{eq:bayes}
\end{equation}
where $\alpha$ reflects the impact of social information (based on the likelihood for $s$),  $\gamma_d$ is the impact of the stimulus presented (based on the likelihood for $d$), and $\beta$ is the bias of the participant towards green (reflecting the Bayesian prior distribution; this analysis could be equivalently stated in terms of bias towards blue). Using this posterior distribution to calculate the probability that each participant chose green, we analyzed a Markov chain on the number of participants in each network choosing ``green'' for each stimulus across waves. In Figure~\ref{fig:statdist} we show that the stationary distribution of this Markov chain---the outcome of transmission through the social network---exaggerates the impact of both $\gamma_d$ and $\beta$, with participants in the networks having more skewed distributions of responses than asocial participants. This model therefore predicts that participating in a social network can amplify the perceptual signal, potentially increasing accuracy, in addition to amplifying people's biases. The simultaneous increase in these two quantities is possible as social information effectively amplifies the signal-to-noise ratio of the stimuli as well as the tendency to make bias-aligned judgments.

\section*{Results}

Experiment 1 was designed to test the prediction that social networks amplify biases. We then developed a simple mitigation method based on our mathematical analysis, which we evaluated in Experiment 2. 

\subsection*{Experiment 1: Bias amplification}
\begin{figure}[t!]
    \centering
    \includegraphics[width=\textwidth]{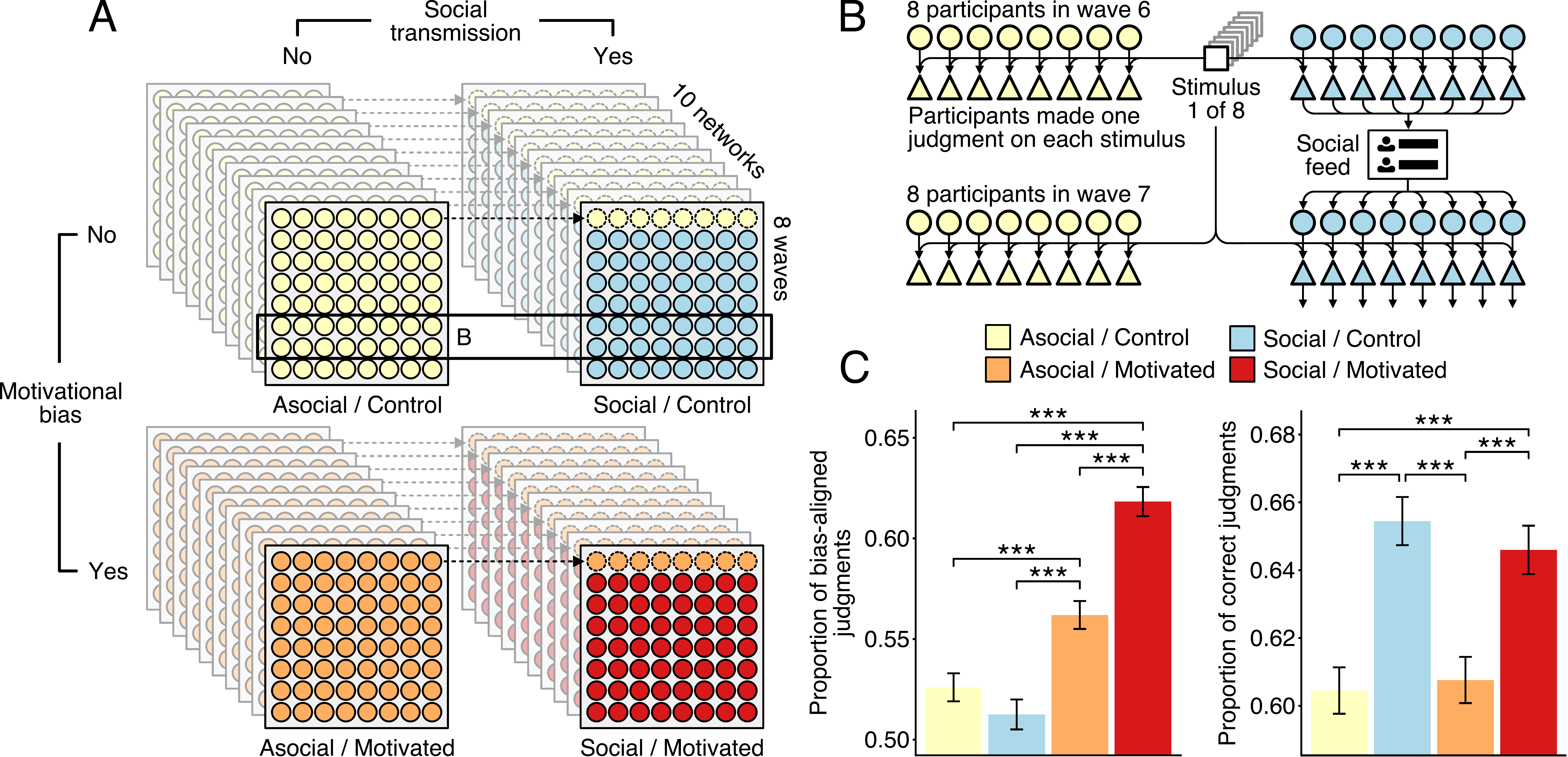}
    \caption{\textbf{Experiment 1 details and results.} (A) Experiment 1 used a 2$\times$2 factorial design, where we varied the presence of social information and of the induced bias in the form of motivated perception. For each condition, we recruited 10 independent networks of 64 (asocial) or 56 (social) participants organized into 8 discrete waves. Dotted line arrows indicates a yoking structure we used to reduce variability by controlling the initial settings across the social and asocial conditions (see Materials and Methods, Experiment 1; Figure~\ref{fig:e1_color_swapping}). (B) Experiment 1 transmission structure in the asocial and social conditions. Circles indicate participants, triangles indicate judgments, and squares indicate stimuli. Each stimulus was defined by a number of motivated color and non-motivated color dots (Materials and Methods, Experiment 1). The judgments made by social participants at each wave were transmitted to the participants in the same network and condition at the next wave. Asocial participants were recruited in waves to facilitate randomization but did not observe any social information. (C) Experiment 1 results averaged across waves and networks for each condition. The plot on the left shows the proportion of trials where participants' judgments corresponded to their motivated color. An unbiased participant would choose their motivated color on 50\% of trials on average. The plot on the right shows the proportion of trials where participants correctly identified the majority color of the dots. For both plots, error bars show standard errors. Stars indicate significance based on a likelihood-ratio test between regressions comparing the relevant conditions, with $^{*}p<0.05$, $^{**}p<0.01$, and $^{***}p<0.001$.
    }
    \label{fig:e1_full}%
\end{figure}

The design and results of our first experiment are shown in Figure~\ref{fig:e1_full}.
We used a factorial design to independently manipulate two factors: the presence or absence of the motivated perception intervention (``motivated'' conditions vs. ``control'' conditions); and whether participants were shown the judgments of other people or not (``social'' conditions vs. ``asocial'' conditions). Combining these manipulations led to four conditions: Asocial/Motivated; Asocial/Control; Social/Motivated; and Social/Control. Participants in each wave were assigned uniformly at random to one of these four conditions. Ten independent networks were recruited for each condition, and 64 participants were assigned to each network (8 waves of 8 participants). Altogether, we recruited 2,400 participants from Amazon Mechanical Turk. 

We designed the task so that motivated perception would not necessarily impact people's ability to perform well: participants made judgments on four stimuli where the correct answer was green, and four where it was blue (trial order was randomized). This design led to a sample size of 19,200 judgments: 5,120 for each asocial condition and 4,480 for each social condition. Fewer participants were assigned to the social conditions because all participants in the first wave were assigned to an asocial condition (see Materials and Methods, Experiment 1). 

To determine whether accuracy and/or bias differed significantly between conditions, we performed a likelihood ratio test between a logistic regression with fixed effects for all conditions and a logistic regression in which the two conditions were coded as a single condition. All statistical models also included random intercepts for each network. Unless otherwise noted, all analyses and statistical models were preregistered before collecting human data. Our preregistered analyses included participant random effects, which we omitted as they led to singular fits. Using the preregistered models does not change any of our findings (see Table~\ref{table:e1_results}).

As predicted, people in social networks made more accurate judgments. Participants in the Social/Control condition identified the correct majority color on 65.4
\unskip\% of trials, significantly more often than both Asocial/Control (accuracy: 60.4
\unskip\%, $\chi^2(1)=25.65$, $p<0.001$
\unskip) and Asocial/Motivated (accuracy: 60.8
\unskip\%, $\chi^2(1)=22.58$, $p<0.001$
\unskip). This advantage was also observed in networks in the Social/Motivated condition, as the model predicted. Social/Motivated participants chose correctly on 64.6
\unskip\% of trials, statistically significantly more often than both Asocial/Control participants ($\chi^2(1)=17.6$, $p<0.001$
\unskip) and Asocial/Motivated participants ($\chi^2(1)=15.07$, $p<0.001$
\unskip). 
Also as predicted, the color-based bonus influenced people's judgments. Participants assigned to the Asocial/Motivated group made significantly more biased judgments (skewed towards their motivated color) than Asocial/Control and Social/Control participants. Asocial/Motivated participants chose their motivated color on 56.2
\unskip\% of trials, compared to 52.6
\unskip\% for Asocial/Control participants ($\chi^2(1)=13.36$, $p<0.001$
\unskip) and 51.2
\unskip\% for Social/Control participants ($\chi^2(1)=23.53$, $p<0.001$
\unskip).

Finally, as predicted, motivated perception was amplified among participants who were part of a social network. Social/Motivated participants chose their motivated color on 61.8
\unskip\% of trials, statistically significantly more often than Asocial/Motivated participants ($\chi^2(1)=31.47$, $p<0.001$
\unskip). In the Supporting Information we report exploratory analyses of the mechanisms underpinning these results, showing that information sharing in motivated networks helped people make more accurate judgments specifically during trials that aligned with their biases, and that social influence cannot be reduced to a simpler perceptual priming mechanism (SI, Exploratory analyses).  

The Bayesian model presented in Equation \ref{eq:bayes} can be directly fit to our behavioral data. We estimated $\alpha$, $\gamma_d$, and $\beta$
using psychometric modeling. This approach draws on techniques developed in the education literature that exploit the relational structure of testing data (each participant evaluated multiple stimuli, and each stimulus was evaluated by multiple participants) to fit a logistic model with separate effects for participants and stimuli \supercite{embretson2013item} (see Materials and Methods, Psychometric model). This allowed us to estimate the degree of bias evident in each participant's decision-making while controlling for the ambiguity of each stimulus and the observed social information.

We used this approach to analyze the judgments made by Asocial/Motivated and Social/Motivated participants (Materials and Methods, Psychometric model). In both models, the mean bias towards motivated color ($\beta$ for participants paid for green dots, $-\beta$ for participants paid for blue dots) was positive and significant, indicating that participants' judgments were skewed in the predicted way (Asocial/Motivated bias mean: 0.26
\unskip, 90\% HDI: [0.219
\unskip, 0.303
\unskip]; Social/Motivated bias mean: 0.371
\unskip, 90\% HDI: [0.302
\unskip, 0.441
\unskip]; full results in Table~\ref{table:e1_irt}). Similarly, there was a monotonic relationship between the proportion of green dots and the estimated coefficient $\gamma_d$ for each stimulus. Finally, in the Social/Motivated model, we observed a positive relationship between the number of prior judgments of green and the probability of choosing green ($\alpha$ = 0.232
\unskip, 90\% HDI: [0.198
\unskip, 0.266
\unskip]). These parameters are predicted to lead to moderately amplified bias (see Figure~\ref{fig:statdist}).

\subsection*{Reducing bias amplification}


Our Bayesian analysis illustrates that bias amplification can arise when people observe judgments produced by others who share the same biases. Bias amplification happens because the bias of one wave is passed forward to the next wave through the sample of observed judgments. This suggests that bias amplification could be reduced by adjusting the transmission process so that people observe a set of judgments that better approximate an unbiased sample, representing a broader range of perspectives. Previous research has sought to construct more representative samples by directly curating opposing viewpoints or through heuristic algorithms that alter people's social connections \supercite{balietti2021reducing,guilbeault2018social,bail2018exposure}. Here, we introduce an algorithmic strategy that makes it possible to obtain more representative samples from within an individual's own network. The general form of this problem---approximating a sample from a target distribution $p$ using samples drawn from a different distribution $q$---can be solved using a method called importance sampling \supercite{tokdar2010importance,elvira2019generalized}. 

Importance sampling provides a simple formula for assigning a weight $w_{jd}$ to each judgment $x_d$ made by individual $j$ in response to stimulus $d$. The weight reflects the relative probability of the judgment under the generating distribution $q_j(x_d)$ specific to participant $j$, and a target sampling distribution $p(x_d)$:
\begin{equation}\label{eqn:weights_formula}
    w_{jd} \propto \frac{p(x_d)}{q_j(x_d)}.
\end{equation}
After computing these weights for the judgments made by every individual in each network, a new sample is drawn from this weighted distribution. Intuitively, this algorithm prioritizes judgments that are underrepresented in each individual's network relative to the target distribution. 

We used the psychometric model described above to calculate the terms in this weighting procedure. Specifically, we calculated $q_j$ using the logistic model in Equation \ref{eq:bayes} with participant-specific bias parameters $\beta$ estimated using hierarchical cognitive modeling at each wave (Materials and Methods, Experiment 2 resampling algorithm). The target sampling distribution $p$ is the posterior distribution implied by setting $\beta=0$, which represents unbiased judgment. This approach is an appropriate model of the broader population in our setting because there were equal numbers of participants with green ($\beta > 0$) and blue ($\beta < 0$) biases, allowing us to assume that the mean of this distribution is zero; in other settings, the target $\beta$ can be set to the empirical population mean (Materials and Methods, Experiment 2 resampling scheme). 

Following this procedure results in an adjustment to the social information observed by participants on each trial (e.g., 6 of 8 people chose green, instead of 7 of 8; see Materials and Methods, Experiment 2 resampling scheme). In more familiar terms, such an adjustment would modify the algorithm used to select the content presented in a social media feed. The feed would still exclusively contain perspectives drawn from each participant's own network, but the algorithm leverages variation within this network to promote perspectives that are more representative of a broader population. Crucially, this algorithm is information-preserving and does not require knowledge or even existence of a ``ground truth.''

\subsection*{Experiment 2: Evaluating resampling}

We hypothesized that our importance sampling strategy would mitigate bias amplification while preserving the benefits of information sharing within our experimental paradigm. To test this hypothesis, we evaluated the efficacy of the strategy in a second large experiment. Using the same experimental paradigm as Experiment 1, we recruited 2,464 participants from Mechanical Turk. We determined this sample size by performing a power analysis on 100 simulations of the experiment. These simulations involved alternatively simulating participants' judgments using a model fit to Experiment 1 and simulating the effects of the resampling algorithm on the simulated data (see Supporting Information, Power analysis).

\begin{figure}[t!]
    \centering
    \includegraphics[width=\textwidth]{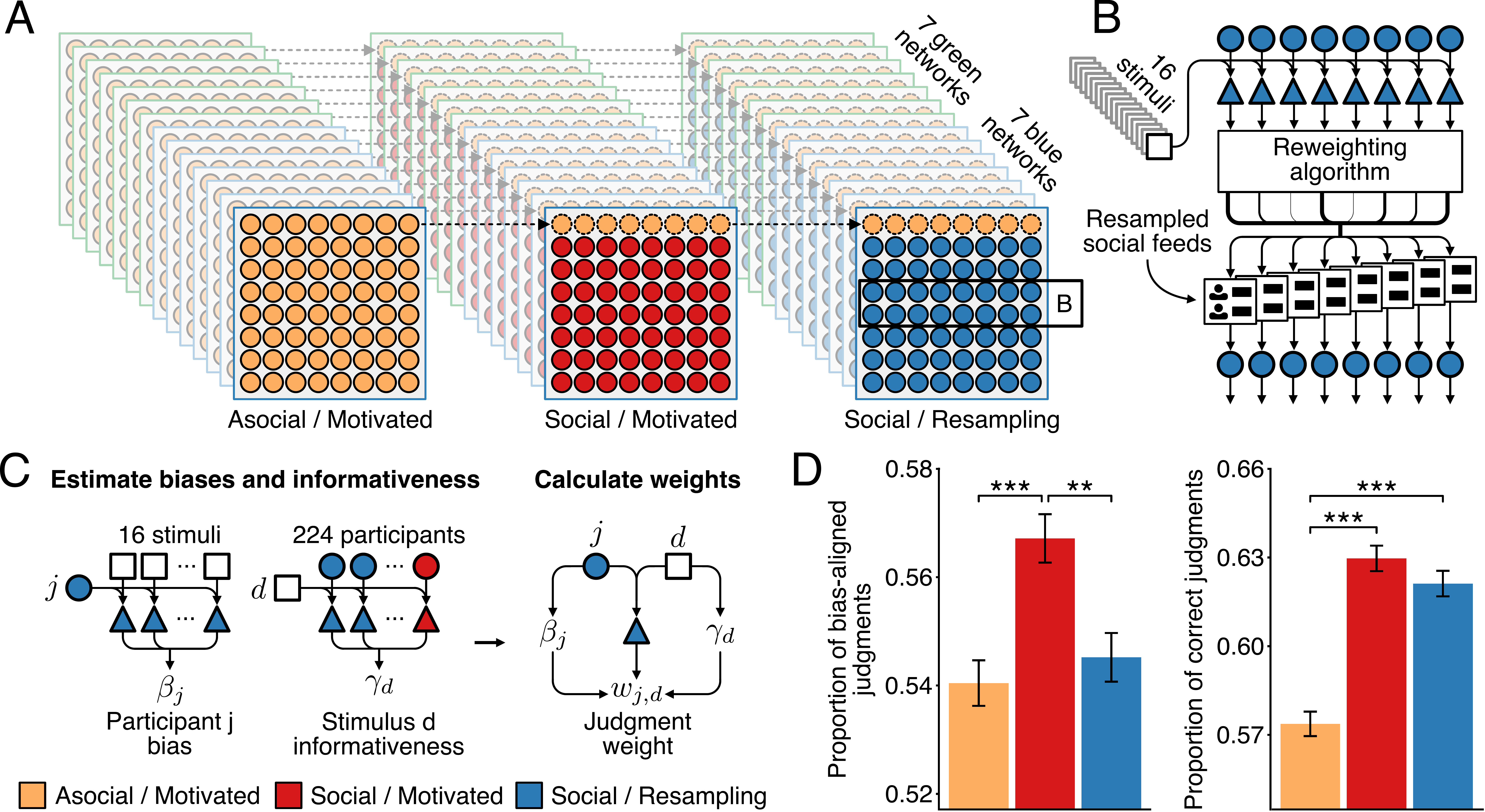}
    \caption{
    \textbf{Experiment 2 details and results.} (A) Experiment 2 had three conditions---Asocial/Motivated, Social/Motivated, and Social/Resampling. For each condition, we recruited 14 independent networks of participants---7 networks of participants paid for green dots, and 7 networks of participants paid for blue dots. As in Experiment 1, participants in each network were organized into 8 discrete waves, with 8 participants in each wave. Dashed lines illustrate the yoking procedure, where second-wave participants in both social conditions observed the judgments made by first-wave Asocial/Motivated participants. (B) The transmission structure for a single network in the Social/Resampling condition. At the end of each wave, we assigned a weight to each judgment made by Social/Resampling participants in that wave. This weight determined the probability of propagating the judgment to participants at the next wave. An independent resampling of judgments was done using the same weights for each participant in the next wave. (C) Calculating the resampling weights for each judgment was a two-step process. We first used a psychometric model to jointly estimate the bias ($\beta$) of each participant $j$ and the ``informativeness'' $\gamma_d$ of each stimulus $d$. This model was fit to Asocial/Motivated participants in wave 1, and to Social/Motivated and Social/Resampling participants in waves 2-8. Biases were estimated from all of a participant's 16 judgments. Informativeness for a stimulus $d$ was estimated using all 224 judgments the model was fit to on that stimulus. We then determined the weight for propagating each judgment using Equation~\ref{eqn:weights_formula}. (D) Experiment 2 results. The plot on the left shows the proportion of trials where participants' judgments corresponded to their motivated color, and the plot on the right the proportion of trials where participants chose correctly. Proportions are averaged across all participants in each condition and error bars show standard errors. Stars indicate statistical significance based on a likelihood-ratio test between models comparing the relevant conditions, with $^{*}p < 0.05$, $^{**}p < 0.01$, and $^{***}p < 0.001$.}
    \label{fig:e2_full}%
\end{figure}

Participants in this experiment were assigned to one of three conditions---Asocial/Motivated, Social/Motivated, and Social/Resampling. The Asocial/Motivated and Social/Motivated conditions replicate Experiment 1. The Social/Resampling condition is analogous to the Social/Motivated condition with the key difference that social information in this condition was subject to the resampling algorithm described above. Participants in social conditions were not informed which social condition they were assigned to. In each condition, half of the networks consisted of participants paid for blue dots (blue ``echo chambers''), and half for green dots (green ``echo chambers''). We recruited 14 independent networks (7 green and 7 blue) per condition, with 64 participants assigned to each network. All participants in the first wave were assigned to the Asocial/Motivated condition. Each participant completed 16 judgments, leading to a sample size of 39,424 judgments: 14,336 for the asocial condition, and 12,544 for each social condition.

We preregistered our main analyses and statistical models before collecting data. To test for differences between two conditions, we performed a likelihood ratio test between an unrestricted regression model with fixed effects for all conditions, and a restricted regression model with the two conditions coded as a single condition. All regression models included a fixed effect capturing the participant's motivated color and random intercepts for each participant.

Replicating Experiment 1, participants in social networks exhibited increases in both bias and accuracy (see Figure~\ref{fig:e2_full}; full results in Table~\ref{table:e2_results}). Asocial/Motivated participants chose their motivated color on 54
\unskip\% of trials, compared to 56.7
\unskip\% for Social/Motivated participants ($\chi^2(1)=12.53$, $p<0.001$
\unskip). Similarly, Asocial/Motivated participants chose correctly on 57.4
\unskip\% of trials, significantly less often than Social/Motivated participants, who chose correctly on 63
\unskip\% of trials ($\chi^2(1)=82.54$, $p<0.001$
\unskip).

As predicted, however, the resampling adjustment mitigated bias amplification. Social/Resampling participants chose their motivated color on 54.5
\unskip\% of trials, a significantly lower rate than Social/Motivated participants ($\chi^2(1)=7.47$, $p=0.006$
\unskip). Furthermore, despite not having access to the true answer for any stimulus, the resampling algorithm preserved the accuracy benefit of social networks. Social/Resampling participants chose correctly on 62.1
\unskip\% of trials, significantly more often than Asocial/Motivated participants ($\chi^2(1)=59.33$, $p<0.001$
\unskip), and not statistically significantly different from Social/Motivated networks ($\chi^2(1)=1.87$, $p=0.172$
\unskip). In the Supporting Information, we report exploratory analyses showing that our algorithm reduced the bias of observed social information in both blue- and green- biased networks, facilitating greater consensus (Figure~\ref{fig:participant_observations}; Supporting Information, Experiment 2 exploratory analyses). The algorithm did so while propagating judgments from each participant at similar rates, rather than e.g. only transmitting judgments made by participants with low estimated bias (Figure~\ref{fig:participant_observations}).

\section*{Discussion}
Together, these results help to identify the mechanisms contributing to bias amplification in social networks. This study recreated a simplified version of the process in a controlled laboratory experiment where we could directly measure the effects of information sharing on bias by comparing the judgments made by isolated and connected individuals. Experiment 1 showed that information sharing led to the amplification of a decision-making bias beyond the levels of bias expressed by individuals completing the task alone. Experiment 2 showed that this amplification effect can be mitigated by a statistical resampling strategy applied to the transmission process. 

The methods we have presented here---using psychometric models to estimate bias and provide a quantitative foundation for resampling---can be extended to more naturalistic settings where the relative probability of each response under a personalized generating distribution and a target sampling distribution can be evaluated. For example, if links that are shared by users of a social media platform can be identified as having a liberal or conservative bias, this information can be used to estimate the biases of individuals who choose to share those links. Evaluating this approach in naturalistic settings is an exciting direction for future research, complementing the detailed causal and theoretical analysis that our laboratory-based approach makes possible.



Online social networks are often constructed by opaque algorithms that prioritize user engagement \supercite{devito2017editors, lazer2015rise}. We have identified and evaluated a simple adjustment to social networking algorithms that reduced bias amplification while maintaining the benefits of information sharing. Resampling allowed participants to observe social information that was generated by people in their network but that better reflected the broader population. This method complements interventions such as fact-checking \supercite{walter2020fact}, education initiatives \supercite{guess2020digital}, and content moderation \supercite{grimmelmann2015virtues}, providing a framework for reducing bias amplification that is transparent, scalable, and underpinned by a mathematical theory that shows why the approach is effective.

\section*{Materials and methods}

\subsection*{Bayesian model}

Our Bayesian model assumes that participants face a decision between two hypotheses: more green dots ($h=g$) or more blue dots ($h=b$). The evidence on which this judgment is based is the observed data $d$ and, optionally, social information $s$ in which $k$ of $n$ other participants endorse green. We thus want to compute $p(h=g|d,s)$. With just two hypotheses, we can write Bayes' rule in log-odds form:
\begin{equation}
\log \frac{p(h=g|d,s)}{p(h=b|d,s)} = \log \frac{p(s|h=g)}{p(s|h=b)} + \log \frac{p(d|h=g)}{p(d|h=b)} + \log \frac{p(h=g)}{p(h=b)} \label{eq:logbayes}
\end{equation}
where we assume that the data $d$ and social information $s$ are independent, with $p(d,s|h) = p(d|h)p(s|h)$, and $p(h)$ reflects prior biases.

For the social information $s$, we assume that each of the judgments is generated independently with probability $1-\epsilon$ of matching the true color. Social information in which $k$ of $n$ judgments favor green thus results in 
\begin{eqnarray}
\frac{p(s|h=g)}{p(s|h=b)} & = &  \frac{(1-\epsilon)^k\epsilon^{n-k}}{\epsilon^k(1-\epsilon)^{n-k}} \\
& = & \left ( \frac{\epsilon}{1-\epsilon} \right )^n \frac{(1-\epsilon)^k\epsilon^{-k}}{\epsilon^k(1-\epsilon)^{-k}} \\
& = & \left ( \frac{\epsilon}{1-\epsilon} \right )^n \left ( \frac{1-\epsilon}{\epsilon} \right )^{2k} \\
& = & \left ( \frac{1-\epsilon}{\epsilon} \right )^{2(k-n/2)}
\end{eqnarray}
which allows us to write
$$
\log \frac{p(s|h=g)}{p(s|h=b)} = (k-n/2) \log \left ( \frac{1-\epsilon}{\epsilon} \right )^{2} .
$$

\noindent Using this result, we can rewrite Equation \ref{eq:logbayes} as
$$
\log \frac{p(h=g|d,s)}{p(h=b|d,s)} = \alpha(k-n/2) + \gamma_d + \beta,
$$
where
\begin{eqnarray}
\alpha & =& \log \left ( \frac{1-\epsilon}{\epsilon} \right )^{2} \\
\gamma_d & = & \log \frac{p(d|h=g)}{p(d|h=b)}\\
\beta & = & \log \frac{p(h=g)}{p(h=b)} .
\end{eqnarray}
The logistic expression given in Equation~\ref{eq:bayes} in the main text results from observing that 
$$
p = \frac{1}{1+\exp \{ - \log \frac{p}{1-p} \}}
$$
and substituting $p(h=g|d,s)$ for $p$.

To model social network dynamics, we define a Markov chain on the number of people endorsing green at time $t$, $k_t$, for a specific stimulus $d$. The transition matrix of the Markov chain is specified by
$$
k_t \sim \mbox{Binomial}\left (n,\frac{1}{1+\exp \{ -\alpha(k_{t-1}-n/2)-\gamma_d-\beta\} } \right ).
$$
Since $\gamma_d$ and $\beta$ behave equivalently in this model, it is sufficient to examine how the stationary distribution of the Markov chain is affected by $\beta$. Figure~\ref{fig:statdist} shows the stationary distribution as a function of $\beta$ for three different values of $\alpha$, reflecting different levels of impact of social information. At all levels of $\alpha$, the stationary distribution amplifies the effect of $\beta$ (and hence $\gamma_d$), with this phenomenon increasing as $\alpha$ increases. Intuitively, this is a form of compounding: $\beta$ and $\gamma_d$ have an effect on judgments at each wave, and the social information conveys that effect to the next wave where it has a further opportunity to influence people's judgments. In the case of $\gamma_d$ this potentially increases accuracy, as people make better use of limited information -- a positive effect of social networks. However, for $\beta$ this is simply a magnification of bias.

\subsection*{Experiment 1}

\subsubsection*{Participants}

We recruited 2,619 participants from Amazon's Mechanical Turk, limiting our study to those located in the United States. We only recruited participants with a Mechanical Turk approval rating of 95\% or higher, and participants could only take the experiment once. Individuals that participated in a pilot version of the study were excluded from the experiment. Participants received a base payment of \$0.65 for completing the experiment, plus an average bonus of \$0.65. Our recruitment algorithm used planned over-recruitment to facilitate efficient data collection in the context of sequential batch recruitment, enabling us to recruit the target sample size of 2,400 participants. Over-recruited participants were paid using the same procedure, but were excluded from analyses and did not have their judgments transmitted to the next wave.

\subsubsection*{Conditions}

Our experiment used a 2$\times$2 factorial design with two binary factors; presence of motivation bias (induced by the experimenters) and availability of social information. Each participant was randomly assigned to one of the four conditions implied by this cross.

The reward structure factor had two levels: motivated and control. In motivated conditions, participants were randomly assigned a motivated color of either green or blue. These participants received one point for every dot of their motivated color on each trial. To ensure the expected total reward for participants in all conditions was equal, participants in control conditions received an additional 400-point reward for completing the experiment. Participants in all conditions also received a 50 point reward on each trial if their judgment was correct. At the end of the experiment, participants' bonuses were paid to them with 10 points equal to one cent.

The social factor also had two levels: asocial and social. Participants in the social conditions observed a set of other participants' responses to a comparable stimulus before each trial. Participants in asocial conditions performed each trial individually. 

\subsubsection*{Wave structure}

Participants were recruited in discrete batches, or ``waves''. A wave of participants was not recruited until all the participants in the previous wave completed the experiment. All participants in the first wave were randomly assigned to one of the two asocial conditions, and participants in later waves were randomly assigned to one of the four conditions.

\subsubsection*{Network structure}

We recruited 10 independent replications (i.e., networks) for each condition. After being assigned to a condition, each participant was then randomly assigned to a replication. Each replication had eight participants at each wave. Replication determined the responses participants in a given wave observed. That is, the replication captured the social network the participant was assigned to. Replication was constant across trials, so a participant observed the responses by the same set of participants on each trial. Participants in asocial conditions were assigned to a replication but did not observe any social information.

Social conditions were yoked to the first-wave asocial replications with the same motivated condition. Social/Motivated participants in the second wave viewed the responses of Asocial/Motivated participants in the first wave and the same replication. Social/Control participants in the second wave viewed the responses of first-wave Asocial/Control participants in the same replication. Participants in later waves in social conditions observed the responses made by participants in the previous wave and same condition and replication. 

We created networks in which participants were rewarded for different colors. To do so, each participant was assigned a marked color of either green or blue. Equal numbers of participants in each network and condition were assigned each color. In the motivated conditions, marked color corresponded to the participant's motivated color. Participants in control conditions, however, were not aware of their marked color. Response data was coded as whether the participant chose their marked color, and the judgments observed by a participant were recoded in terms of that participant's marked color (see Figure~\ref{fig:e1_color_swapping}).

\subsubsection*{Procedure}

All practice and test stimuli consisted of 100 randomly positioned blue and green dots displayed for 1000 milliseconds. After viewing a stimulus, participants judged whether the image contained more blue or more green dots. The blue (\#007ef8, LAB: 53.49, 16.3, -69.16) and green (\#009500, LAB: 53.49, -57.73, 55.72) colors were chosen for identical lightness in LAB color space. A fixation cross and bounding box preceded presentation of the stimuli for 600 milliseconds.

For each stimulus, the position and size of dots varied between replications and waves, but not across conditions. That is, for each replication, participants in condition A observed with the exact stimuli viewed by participants in condition B. For example, all participants in wave $t$ and replication $r$ observed the exact same display ($S_{t,r}$) for each stimulus. Participants in the next wave but same replication making judgments on the same stimulus observed a display ($S_{t+1, r}$) with the same number of marked color dots as $S_{t,r}$, but which were configured in different sizes and positions. A participant's marked color then determined the color of each dot.

Each participant first completed two practice trials. Participants received accuracy and reward feedback after each practice trial. Participants were then required to pass a comprehension test in their first three tries to complete the experiment. Each participant completed 8 test trials. No trial-by-trial feedback was given on test trials. Instead, participants were told their earnings and performance at the end of the experiment. 

Each practice and test stimulus was associated with a number of marked and non-marked color dots. Practice stimuli had 47 and 53 marked-color dots. Test stimuli had 48, 49, 51, or 52 marked-color dots, with two trials for each unique number. Trial order for both practice and test trials was randomized.

Social information was presented as the number of participant making the majority judgment. For example, if 6 of 8 participants in the previous wave chose their marked color, then a participant $P$ at the next wave whose marked color was blue was informed that 6 of 8 participants in the previous wave chose blue. However, if only 3 of 8 participants chose their marked color, then $P$ was informed that 5 of 8 participants chose green. When equal numbers of participants chose their marked and non-marked colors, we used simple randomization to determine which color was presented as the response of 4 out of 8 individuals. The social icons observed before the stimulus were presented again above the blue and green response buttons when participants made their judgment.

We used cover stories to make the task more intuitive. All participants were informed that they were working for an imaginary mining company looking for valuable gemstones. Participants in motivated conditions whose marked color was blue were told that they would be looking for blue sapphires in green grass, and to inspect areas and judge if the area had more sapphire dots or more grass dots. Participants in motivated conditions whose marked color was blue were given a similar cover story, but were instructed that they were looking for green emeralds in blue water. Participants in control conditions were told to judge whether the area contained more blue sapphires or green emeralds.

\subsection*{Psychometric model}\label{e1_irt_model}

We fit separate versions of the Bayesian model introduced above to the judgments made by Asocial/Motivated and Social/Motivated participants in Experiment 1. These models can also be interpreted in terms of Item Response Theory \supercite{embretson2013item}, a psychometric approach typically used to jointly estimate individuals' abilities and questions' difficulties using testing data. We used this approach to estimate the impact of social information $\alpha$, the informativeness of the data $\gamma_d$ and the biases of participants $\beta$. 

The model specified in Equation~\ref{eq:bayes} in the main paper is a form of logistic regression. We fit two models---one for the Social/Motivated condition and one for the Asocial/Motivated condition---via Bayesian logistic regression. 
We estimated the bias $b_j$ of participant $j$ towards their motivated color using a hierarchical Bayesian model, with $b_j \sim \mathcal{N}(\mu_b,\sigma_b)$, $\mu_b \sim \mathcal{N}(0,3)$, and $\sigma_b \sim \text{Lognormal}(0,2)$. This model allows us to pool data across participant groups and increases statistical power. The estimate $\mu_b$ of the mean of the population-level distribution captures the tendency of blue- and green-biased participants to choose their motivated color. This can be then recoded as the bias $\beta_j$ of participant $j$ towards green in the regression model:
\begin{equation}
    \beta_j = 
    \begin{cases}
        b_j, & \text{if } j \text{'s motivated color is green}\\
        -b_j, & \text{if } j \text{'s motivated color is blue.}\\
    \end{cases}
\end{equation}

To model each stimulus, both models included coefficients $\boldsymbol{\gamma_d} \sim \mathcal{N}(0,20)$ on each level of a dummy variable representing the number of green dots (0.48, 0.49, 0.51, 0.52) in the stimulus. We modeled the impact of the social information $s$ on Social/Motivated participants' judgments by fitting a weight $\alpha \sim \mathcal{N}(0,20)$ on $k-n/2$, or the number of green judgments observed for the stimulus minus four ($n/2$). The asocial model did not include $\alpha$. Controlling for this social information allowed us to compare the individual-level bias of Asocial/Motivated and Social/Motivated participants using the bias term $\mu_b$ as described above.

Parameters in both models were estimated from 8 Markov chain Monte Carlo chains run for 2,500 iterations using the No-U-Turn variant of Hamiltonian Monte Carlo \supercite{hoffman2014no}. The first 1,250 iterations of each chain were discarded and not used for parameter estimation. Average sampled values and highest density intervals for both models are given in Table~\ref{table:e1_irt}.

\subsection*{Experiment 2 resampling algorithm}

The resampling algorithm used in Experiment 2 involved (1) fitting a psychometric model to estimate participant biases and stimulus informativeness, (2) estimating weights for each judgment using multiple importance sampling, and (3) resampling a set of judgments to propagate to participants at the next wave. Here, we describe each step of this process in detail.

\subsubsection*{Estimating $\beta$ and $\gamma_d$}


We modeled the generating distribution $q$ and the unbiased target distribution $p$ using a psychometric model based on the Bayesian model introduced above. 
Participant biases were assumed to be $\beta_j \sim N(0,3)$ and stimulus informativeness was $\gamma_d \sim N(0,3)$. We omitted $\alpha$ from this version of the model. We did this to establish that the resampling algorithm can be applied in the most conservative case where the observation history of the people whose judgments are being resampled is unknown. To estimate $q$ we use a Bayesian logistic regression model that incorporates both of these factors. To estimate $p$, we calculate the distribution that results from setting $\beta_j=0$. Unlike the Experiment 1 psychometric model, coefficients were estimated for each stimulus directly from judgment data. That is, we fit $\gamma_d$ without using the true dot proportions (i.e., the correct answer). Additionally, our approach does not require knowledge of the true bias of any individual. Instead, our framework allows us to estimate biases through distributional assumptions about the population.

At the end of each wave, we fit the IRT model by running 8 Markov chain Monte Carlo chains for 2,500 iterations (with the first half warmup) using the No-U-Turn variant of Hamiltonian Monte Carlo \supercite{hoffman2014no}. In the first wave, the model was fit to Asocial/Motivated participants, and in later waves it was fit to Social/Motivated and Social/Resampling participants.

\subsubsection*{Multiple importance sampling}

We aimed to obtain an unbiased sample of judgments while only having access to judgments produced by a biased population. This problem---approximating a sample from a target distribution $p$ using samples drawn from a different distribution $q$---can be solved with importance sampling \supercite{tokdar2010importance}. In importance sampling, a weight $\tilde{w}_i$ is assigned to each sample $x_i$, with:

\begin{equation}\label{eqn:importance-sampling-vanilla}
\tilde{w}_i = \frac{p(x_i)}{q(x_i)}.
\end{equation}

\noindent This setup assumes that all observations are drawn from a common generating distribution $q$. In our setting, however, bias can vary across individuals. This means that the observed judgments are drawn a distinct generating distribution $q_j$ for each participant $j$. This requires a more specific form of adjustment known as multiple importance sampling \supercite{elvira2019generalized}.

One method for performing multiple importance sampling is to set the weight of each observation equal to its probability under $p$ divided by its probability under $q_j$: 
\begin{equation}\label{eqn:multiple-importance-sampling}
\tilde{w}_{i} = \frac{p(x_i)}{q_j(x_{i})}.
\end{equation}
The weights are then normalized so they can be interpreted as components of a probability vector. This can be done by dividing each weight by the sum of the weights for all $N$ observations:

\begin{equation}\label{eqn:importance-normalization}
w_{i} = \frac{\tilde{w}_{i}}{\sum_{n=1}^{N}\tilde{w}_n}.
\end{equation}

\subsubsection*{Resampling judgments}

In our setting, $q$ is the distribution of biased judgments and $p$ is the unbiased distribution. Given a sample from $q$, we can use importance sampling to draw a sample of judgments that is closer to $p$. As described above, this procedure involves assigning weights to the original set of judgments and then sampling a new set of judgments from the set with probabilities proportional to the weights.

At the end of each wave, we used the estimated IRT models to calculate $p$, the probability of the judgment under an average bias towards green, and $q$, the probability of the judgment given the participant's bias for all judgments in the wave. We used Equation~\ref{eqn:multiple-importance-sampling} to calculate the unnormalized weight $\tilde{w}_{jd}$ for every judgment made by Asocial/Motivated (wave 1) or Social/Resampling (waves 2-8) participants. To normalize the weights, we assigned each judgment to a judgment set. A judgment set consisted of the $N=8$ judgments made by the participants in the same network and wave on stimulus $d$. Following Equation~\ref{eqn:importance-normalization}, each weight was normalized by dividing it by the sum of all the weights in the judgment set.

The normalized weights $\boldsymbol{w}$ determined the probabilities of propagating each judgment to participants at the next wave. On every Social/Resampling trial, we used these probabilities to sample a set of eight judgments from the relevant judgment set. Judgments were resampled with replacement, and distinct samples were propagated to each participant at the next wave.

\subsection*{Experiment 2}\label{appendix:e2_methods}

\subsubsection*{Participants}

We recruited 2,518 participants from Amazon's Mechanical Turk. We required participants to be based in the United States, have an approval rating of at least 95\%, and to have successfully completed at least 5,000 HITs. Participants could only take the experiment once. We allowed individuals who completed Experiment 1 or its pilot studies to participate in the experiment. However, individuals that participated in a pilot version of Experiment 2 were excluded. Participants received a base payment of \$1.20 plus an average bonus of \$1.28. As in Experiment 1, we used an over-recruitment algorithm to accelerate collection of our target sample size of 2,464 participants. Over-recruited participants were paid using the same procedure, but were excluded from analyses and did have their judgments transmitted to the next wave.

\subsubsection*{Conditions}

All participants were assigned a motivated color for which received a one point bonus for every dot of that color. In addition to their color bonus, all participants received a 50 point bonus on each trial if their judgment was correct. At the end of the experiment, participants' bonuses were paid to them with 10 points equal to one cent.

Participants were assigned to one of three conditions\textemdash Asocial/Motivated, Social/Motivated, and Social/Resampling. Participants in the asocial condition completed the experiment individually and did not observe social information. Social participants observed a set of judgments made by participants in the previous wave on the same stimulus before making a judgment. Social/Motivated participants observed the original judgment set. Participants in Social/Resampling networks observed a set of 8 judgments sampled with replacement from the judgment set. We resampled a separate set of 8 judgments for each participant.

\subsubsection*{Wave structure}

As in Experiment 1, participants were recruited in 8 discrete waves. Networks and conditions were run in parallel, so participants at wave $t+1$ were not recruited until all participants in wave $t$ completed the experiment. All participants in the first wave were assigned to the asocial condition. Equal numbers of participants in later waves were assigned to each condition using block randomization.

\subsubsection*{Network structure}

At each wave, participants in each condition were randomly assigned to a replication. We recruited 14 replications for each condition. A replication determined the responses social participants observed. Replication was fixed, so a social participant observed social information from the same set of participants on each trial. 

As in Experiment 1, each replication had 8 participants in each wave. Instead of assigning participants with both green and blue biases to the same network as in Experiment 1, networks were composed entirely of participants with the same motivated color. Seven of the replications for each condition consisted of participants biased towards green, with the other seven populated by participants biased towards green. This structure allowed to avoid color-swapping within networks as we did in Experiment 1.

Social/Motivated and Social/Resampling networks were yoked to the same initial asocial wave. That is, two sets of social participants (Social/Motivated and Social/Resampling) in replication $r$ making a judgment on stimulus $s$ in wave 2 observed the judgments made by the same set of Asocial/Motivated participants on stimulus $s$ in wave 1. At later waves $t>2$, social participants in condition $c$, replication $r$, and wave $t$ making a judgment on stimulus $s$ observed the responses made by participants on stimulus $s$ in condition $c$, replication $r$, and wave $t-1$.

\subsubsection*{Procedure}

As in Experiment 1, all stimuli consisted of 100 dots displayed for one second. Participants then judged whether there were more blue or more green dots in the stimulus. The blue (\#007ef8) and green (\#009500) colors matched those used in Experiment 1. A fixation cross and bounding box was displayed for 600 milliseconds prior to presentation of the stimuli. Positioning and sizing for a given stimulus was determined randomly under the constraint that not dots overlap with each other or the stimulus boundary. Unlike Experiment 1, we did not constrain dot positioning and sizing to match on each wave and replication.

Social information on each trial was shown before participants in social conditions viewed a stimulus. Social information consisted of both text and icons, with $k$ icons colored to match the judgment of the majority, and $8-k$ colored in gray. For example, 6 green and 2 blue judgments were displayed as 6 green and 2 gray icons. However, if only 3 of 8 participants chose green, participants were shown 5 blue icons and 3 gray icons. Ties were broken via simple randomization. The same social icons a participant observed before a stimulus were also presented above the blue and green response buttons when participants made their judgment.

Participants first completed two practice trials with 53 and 47 green dots. Participants received accuracy and reward feedback after both practice trials. Practice trials did not count towards participants' bonuses, and no resampling was performed on practice trials in Social/Resampling networks. Participants then completed a comprehension test before starting test rounds. Participants that failed to answer every test question correctly in their first three tries were excluded from the experiment.

As in Experiment 1, test stimuli had either 48, 49, 51, or 52 green dots. Participants completed 16 test trials, with four test trials for every number of green dots. No trial-by-trial feedback was given on the test trials. Instead, participants were informed of their earnings at the end of the experiment. Trial order for both practice and test trials was randomized.

To make the task more intuitive, we used the same cover stories we used in Experiment 2. Participants were told that they were working for an imaginary mining company looking for valuable gemstones. Participants whose motivated color was blue were instructed that they were looking for blue sapphires in green grass, and those whose motivated color was green were told that they were looking for green emeralds in blue water.

\subsection*{Code and data availability}

Code, data, and links to preregistrations are available at https://github.com/mdahardy/bias-amplification. Both experiments were built using \texttt{Dallinger} (Experiment 1: 5.1.0; Experiment 2: custom fork). The resampling algorithm in Experiment 2 was implemented in Python (3.9) using \texttt{NumPyro} (0.7.2). The resampling algorithm for the Experiment 2 power analysis and the psychometric models of Experiment 1 were implemented in R using \texttt{Rstan} \supercite{carpenter2017stan} (2.21.3). Mixed effects models were implemented in R (4.1.2) using \texttt{lme4} (1.1.28).

\printbibliography[title={References}]
\end{refsection}
\begin{refsection}

\renewcommand{\thetable}{S\arabic{table}}
\renewcommand{\thefigure}{S\arabic{figure}}

\setcounter{figure}{0}
\setcounter{table}{0}

\setcounter{equation}{0}

\newpage
\section*{Supporting Information}

\subsection*{Exploratory analyses}

\subsubsection*{Experiment 1 exploratory analyses}

To investigate how social observed amplified both bias and accuracy, we performed exploratory analyses by partitioning the data from Experiment 1 into two subsets based on whether the true answer matched a participant's motivated color. As in the main text, we determined whether accuracy varied between two conditions by performing a likelihood ratio test between a regression where the two conditions were coded separately and one where they were treated to a single condition. All regressions included random intercepts for each replication (see Materials and Methods, Experiment 1), and were only fit on data from the relevant stimulus type (matched or nonmatched).

Among participants in motivated conditions, social observation increased the accuracy of participants' judgments on matched stimuli from 67
\unskip\% to 76.4
\unskip\% ($\chi^2(1)=53.28$, $p<0.001$
\unskip). However, social observation had no effect on motivated participants' accuracy on nonmatched stimuli (Asocial/Motivated accuracy: 54.6
\unskip\%; Social/Motivated accuracy: 52.8
\unskip\%; $\chi^2(1)=1.57$, $p=0.211$
\unskip).

For participants in control networks, however, social observation increased accuracy on both types of stimuli. On matched stimuli, Asocial/Control participants chose correctly on 63
\unskip\% of stimuli, whereas Social/Control participants chose correctly on 66.7
\unskip\% of stimuli ($\chi^2(1)=7.03$, $p=0.008$
\unskip). On unmatched stimuli, social observation increased participants' accuracy from 57.9
\unskip\% to 64.2
\unskip\% ($\chi^2(1)=20.28$, $p<0.001$
\unskip).

\subsubsection*{Perceptual priming}

To investigate whether bias amplification was due to a perceptual priming effect, we performed exploratory analyses on the judgments made by Social/Motivated participants on ``tie'' stimuli, or those when the social information consisted of 4 green and 4 blue judgments. On these stimuli, we used simple randomization to determine whether green or blue was presented as the majority judgment (see Materials Methods, Experiment 1; Materials and Methods, Experiment 2). If social observation primarily influenced participants' judgments through perceptual priming, the color selected as the majority judgment on these stimuli should predict participants' judgments.

In Experiment 1, Social/Motivated participants made judgments on 792
\unskip tie stimuli. We fit a logistic regression model predicting whether participants chose green on these stimuli with independent variables capturing whether the participant's motivated color was green (bias-green), whether green was presented as the majority color (majority-green), and the proportion of green dots. This regression revealed a significant and positive effect of bias-green ($\beta=0.76$, $z=4.89$, $p<0.001$
\unskip) but no effect of majority-green ($\beta=0.05$, $z=0.32$, $p=0.753$
\unskip). We observed similar results on the 3,758
\unskip Social/Motivated and Social/Resampling tie stimuli in Experiment 2. Using the same logistic model we used for Experiment 1, we found that bias-green had a significant and positive effect on the probability of choosing green on tie stimuli ($\beta=0.37$, $z=5.57$, $p<0.001$
\unskip). However, there was no effect of majority-green on participants' judgments ($\beta=0$, $z=0.01$, $p=0.996$
\unskip). These results suggest that bias amplification was primarily driven by the content of the social information, rather than a perceptual priming effect caused by observing green or blue icons before each stimulus

\subsubsection*{Experiment 2 exploratory analyses}

Here, we report exploratory analyses investigating the impact of the resampling algorithm on the social information participants in Experiment 2 observed. Our first set of analyses tested whether the bias and accuracy of the observed social information differed between the social conditions. To do so, we fit unrestricted and restricted mixed-effects regression models predicting both (1) the accuracy and (2) the bias of the observed social information. All models included a fixed effect capturing the participant's motivated color and random intercepts for each replication (see Materials and Methods, Experiment 2). The unrestricted models had an additional fixed effect on the participant's condition. We then determined whether the accuracy or bias differed between the conditions by performing a likelihood ratio test on the relevant unrestricted and restricted models.

We found that resampling significantly reduced bias in the observed information: 56.4
\unskip\% of the social information observed by Social/Motivated participants conformed to their bias, compared to just 50.8
\unskip\% for Social/Resampling participants ($\chi^2(1)=293.27$, $p<0.001$
\unskip). Furthermore, the resampling algorithm had a small but significant effect on the accuracy of social information. Social/Motivated participants observed accurate social information 62.3
\unskip\% of the time, whereas 64
\unskip\% of the social information observed by Social/Resampling participants was accurate ($\chi^2(1)=32.62$, $p<0.001$
\unskip).

We used a similar approach to investigate whether both blue- and green-biased Social/Resampling participants tended to observe less biased social information. To compare the proportion of green judgments observed by blue-biased participants in both conditions, we fit both unrestricted (with a fixed effect for condition) and restricted (no fixed effect for condition) regression models to the judgments made by blue-biased social participants. Both models predicted the proportion of green judgments observed on each trial and included random intercepts for each replication. We then performed a likelihood-ratio test between the unrestricted and restricted models to measure whether the proportion of green judgments observed in both conditions differed. This approach revealed that among blue-biased social participants, Social/Resampling participants observed significantly more green judgments compared to Social/Motivated participants ($\chi^2(1)=183.08$, $p<0.001$
\unskip). Using a similar set of models, we also found that our resampling scheme increased the proportion of blue judgments observed by green-biased social participants ($\chi^2(1)=111.83$, $p<0.001$
\unskip).

\subsection*{Experiment 2 power analysis}

\begin{figure}
\centering
\includegraphics[width=\textwidth]{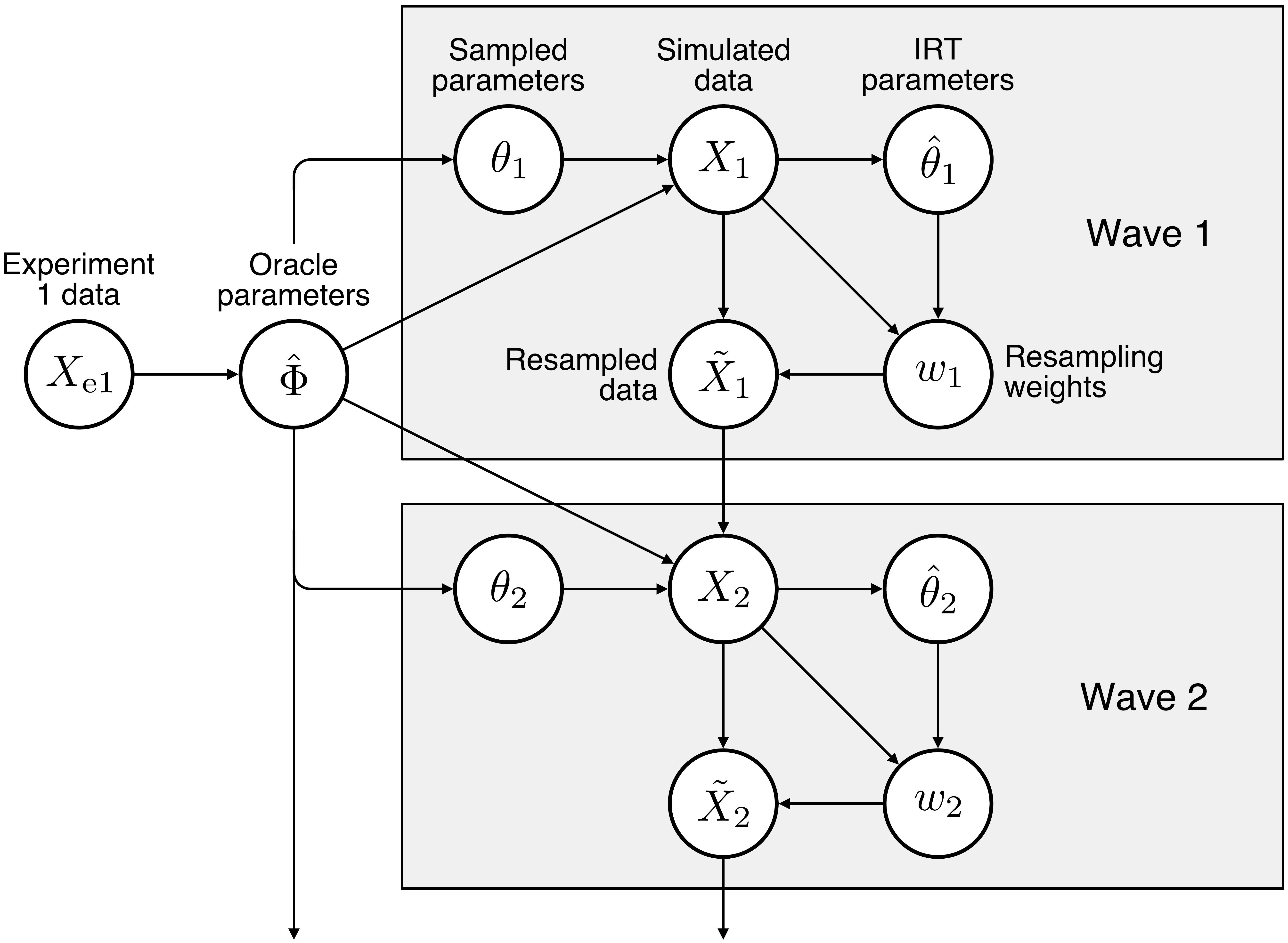}
\caption{\textbf{Simulation design for the first two waves of our Experiment 2 power analysis.} Our design alternated between simulating participants' judgments and the effects of our resampling procedure. We first fit the parameters $\hat{\Phi}$ of the oracle models to Experiment 1 data $X_{e1}$ using Markov Chain Monte Carlo. One model was fit to Asocial/Motivated participants, and one to Social/Motivated participants. At each wave $t$, we used either the asocial (wave 1) or social (waves 2-8) oracle to sample participant biases $\theta_t$ and simulate judgments $X_t$. To simulate our resampling procedure, we then fit IRT parameters $\hat{\theta}_t$ to the simulated judgments at each wave. As in the experiment, these IRT models did not have access to the ground truth or participants' true biases. We used our fitted IRT model to determine the importance weights $w_t$ for each judgment and resample a set of judgments $\tilde{X}_t$ to propagate to the next wave. For each simulation, we repeated this process for 8 waves (the same fitted oracle model was used in all simulations). Stimuli, network structure, and sample sizes matched the setup used in Experiment 2.}
\label{fig:simulation_design}
\end{figure}

Prior to running Experiment 2, we simulated our proposed experiment design and resampling scheme and used this data to perform a power analysis. Simulating the experiment involved simulating both participants' judgments and the effects of our resampling algorithm on these generated judgments (see Figure~\ref{fig:simulation_design}). Importantly, both simulation steps used models fit to different data and had access to different information.

To simulate participants' judgments, we fit oracle models to the judgments made by Asocial/Motivated and Social/Motivated participants in Experiment 1. Both oracle models used a Bayesian logistic regression to predict whether a participant chose green on each stimulus, and included a hierarchical term capturing the bias of each participant towards their motivated color. The prior for the bias $b_j$ for participant $j$ was set to $b_j \sim \mathcal{N}(\mu_b,\sigma_b)$ with $\mu_b \sim \mathcal{N}(0,3)$ and $\sigma_b \sim \text{Lognormal}(0,2)$. Both models also included a constant $c \sim \mathcal{N}(0,20)$, and separate weights $\boldsymbol{\gamma{d}} \sim \mathcal{N}(0,20)$ on each level of an indicator capturing the number of green dots (0.48, 0.49, 0.51, 0.52) on the stimulus. The Social/Motivated model included an additional weight $v \sim \mathcal{N}(0,20)$ on the number of green judgments observed for each stimulus.

To simulate our resampling scheme, we fit a Bayesian IRT model to the simulated data from each wave. As in the experiment (and unlike the oracles models), this IRT model did not have access to the ground truth of any stimulus or participants' biases (see Materials and Methods, Experiment 2 psychometric model). In the first wave, this model was fit to the simulated judgments made by the first Asocial/Motivated wave, and to social participants in later waves. We used this IRT model to determine the resampling weights using the same procedure we used in Experiment 2.

\begin{figure}
\centering
\includegraphics[width=\textwidth]{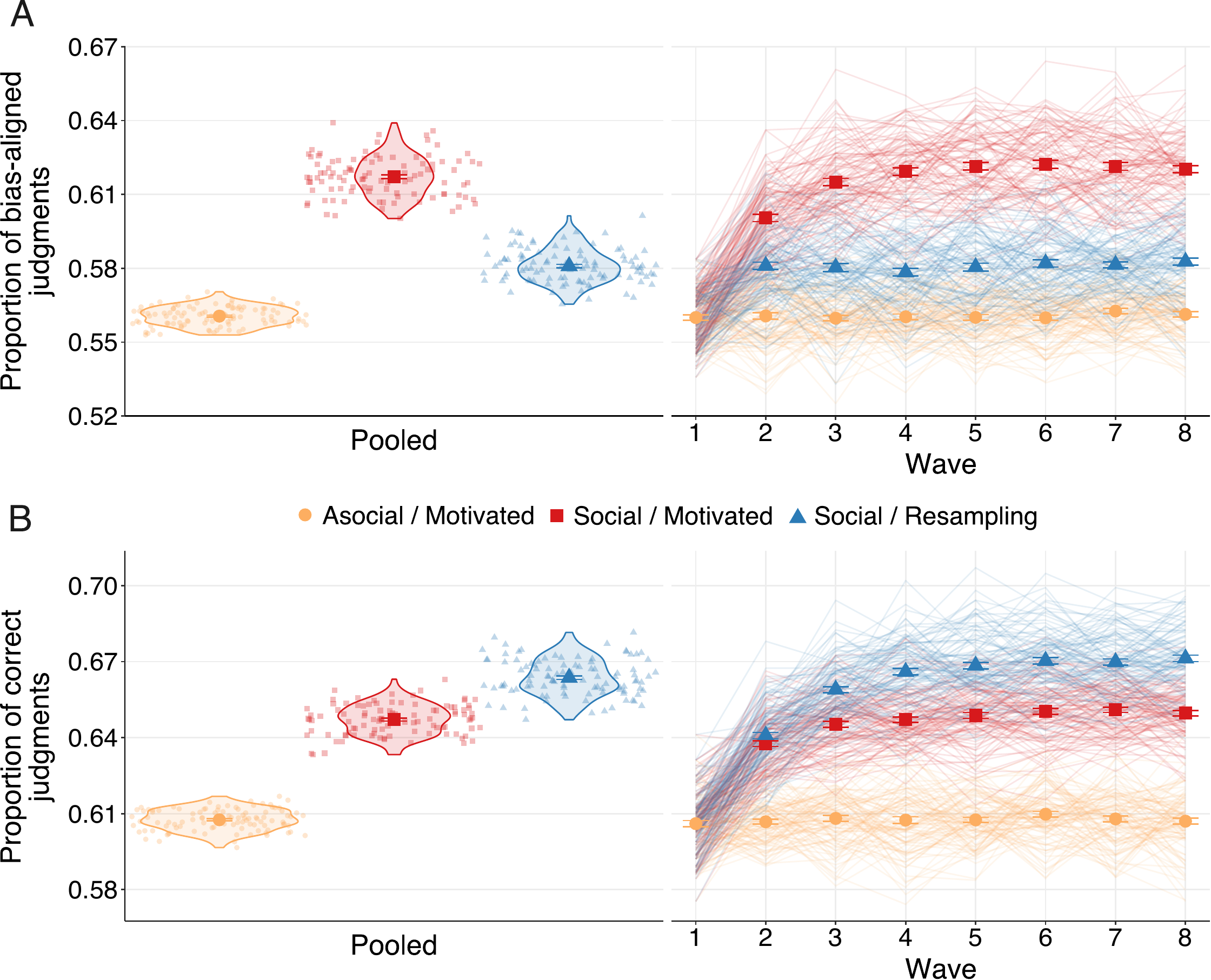}
\caption{\textbf{Simulation results for our Experiment 2 power analysis.} We recorded the proportion of both biased (A) and correct (B) judgments in the simulated data. For both, plots on the left show the results pooled across waves, and those on the right show the results by wave. Large points give the relevant statistic (bias or accuracy) pooled across simulations. Small points in the pooled plots and lines in the wave plots show the results of each simulated experiment (n=100). Error bars show standard errors of the experiment means.}
\label{fig:simulation_results}
\end{figure}

We simulated our proposed Experiment 2 design 100 times (results are shown in Figure~\ref{fig:simulation_results}). Before simulating an experiment, we fit the Asocial/Motivated and Social/Motivated oracle models to data from Experiment 1 using the Stan programming language. To fit these models, we ran 8 MCMC chains for 2,500 iterations, with the first half warmup. The same two oracle models were used for all 100 experiment simulations. For each simulation, we simulated 14 independent networks for each of our three experiment conditions (Asocial/Motivated, Social/Motivated, and Social/Resampling). Within each condition, 7 of the networks were composed of participants biased towards blue, and 7 towards green. As in Experiment 2, each network consisted of 64 participants organized into 8 discrete waves.

At the start of each wave, we first sampled a bias for each participant from the normal distribution defined by the mean oracle bias parameters. We then used this bias and the mean parameter estimates from the oracle models to simulate participants' judgments on each stimulus. Each simulated participant made 16 judgments, with the dot proportions matching those used in the real experiment. Social/Motivated and Social/Resampling participants observed the original (Social/Motivated) or resampled (Social/Resampling) judgments made by simulated participants in the previous wave. Mirroring the yoking structure we used in Experiment 2, social participants in the second wave observed judgments made by first-wave Asocial/Motivated participants in the same replication. Social participants in later waves observed the judgments made by participants in the previous wave and same condition and replication. Asocial/Motivated participants did not observe any social information.

We used the same hypothesis tests and statistical models to perform our power analysis that we used in Experiment 2. We compared conditions using a likelihood ratio test between a model where the two conditions were treated separately, and one where they were coded as the same condition. Both models had fixed effects for motivated color and random intercepts for participants and social networks. However, if either model had singular fit, we removed random terms from both models until there was no singularity. Occasionally, this meant running our hypothesis test using models with only fixed effects. We performed this model search separately for each hypothesis test and simulation. In 99 of the 100 simulated experiments, all of our hypotheses were as expected. We therefore determined that our proposed Experiment 2 design was sufficiently powerful.

\begin{figure}
\centering
\includegraphics[width=\textwidth]{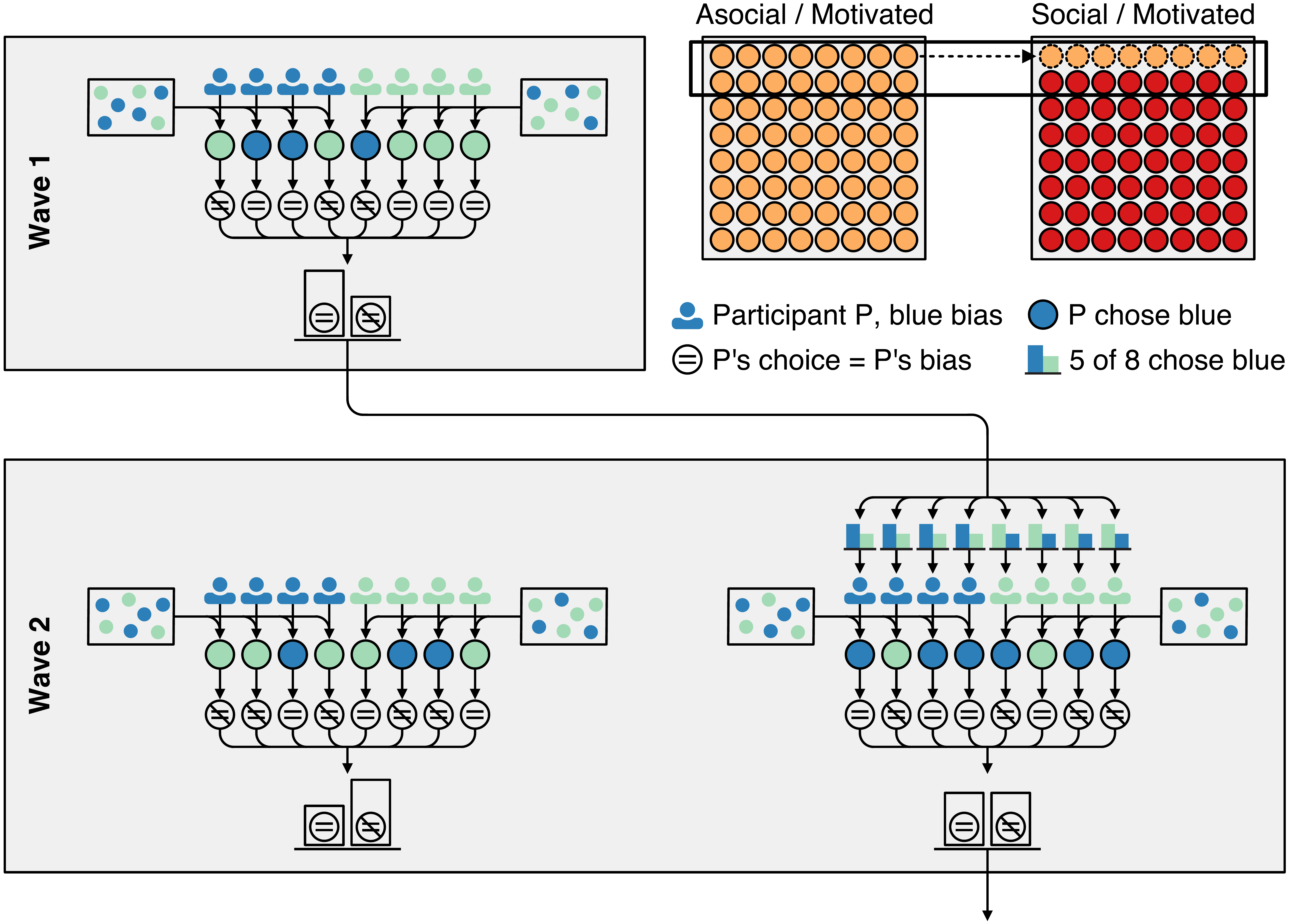}
\caption{\textbf{Experiment 1 yoking structure and marked color design.} All participants in the first wave were assigned to an asocial condition. In the second wave, social participants observed the judgments made by first-wave asocial participants in the same motivated condition and replication (see Materials and Methods, Experiment 1). Each network consisted of four participants with a marked color of blue and four of green at each wave. A participant's marked color determined the color of the dots in the stimulus (dot sizes and positions were fixed in a wave and replication). To match this color-swapping, social information was also presented in terms of the participant's marked color. That is, if 5 of 8 people chose their marked color in the previous wave, participants with a marked color of green would be told that 5 of 8 people chose green, and participants with a marked color of blue that 5 of 8 participants chose blue. Marked color corresponded to motivated color for participants in motivated conditions. Participants in control conditions were assigned marked colors using an identical process, but were not informed of their marked color.}
\label{fig:e1_color_swapping}
\end{figure}

\begin{figure}
\centering
\includegraphics[width=\textwidth]{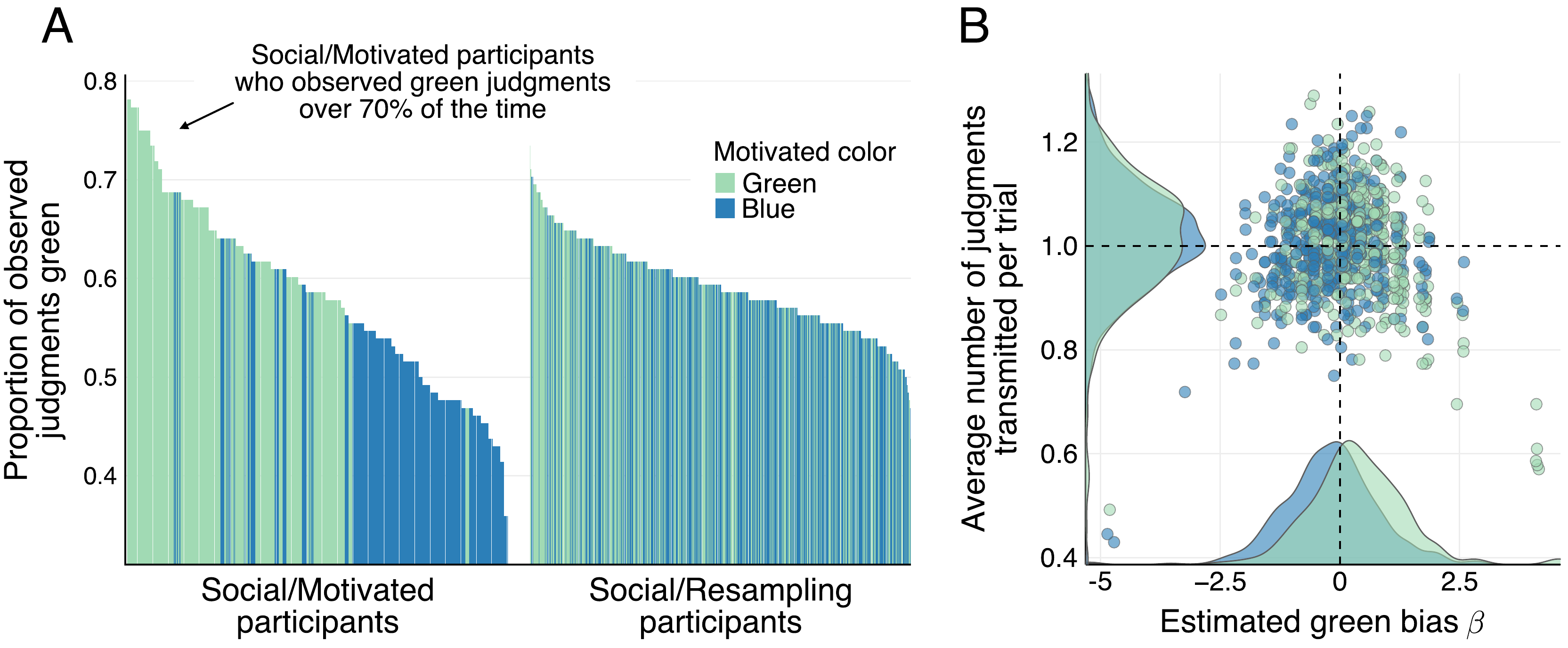}
\caption{\textbf{Experiment 2 participant observations and estimated biases.} (A) The resampling algorithm increased consensus between networks with induced biases towards green and blue. Each bar shows the proportion of green judgments from the previous wave that each participant ($n$=784 for both conditions) observed over all 16 trials of the experiment. Bars are arranged in descending order, and bar color corresponds to the participant's motivated color. (B) Estimated participant biases and transmission rates in the Social/Resampling condition. In our resampling algorithm, each participant's judgment could be propagated multiple times to a participant at the next wave. Rather than only propagating judgments made by those with low estimated bias, the algorithm transmitted each participant's judgments at similar rates. Points show the estimated green biases of participants in the Asocial/Motivated (wave 1) and Social/Motivated (waves 2-7) condition and the number of times their judgments were transmitted.}
\label{fig:participant_observations}
\end{figure}

\begin{table}[p]
\centering
\renewcommand{\arraystretch}{1.2}
\setlength\tabcolsep{3.5pt}
\begin{tabular}{crcccccccc}
& & \multicolumn{2}{c}{\textbf{Asocial / Control}} \\
& & Reported & Prereg 
\\
\cline{1-4}
\multirow{2}{*}{\textbf{\shortstack{Asocial / \\ Motivated}}} 
& Bias & 
$13.4^{***}$
 & 
$12.7^{***}$
 & 
\multicolumn{2}{c}{\textbf{Asocial / Motivated}} & 
\\ 
& Accuracy & 
$0.1$
 &
$0.1$
 &
Reported & Prereg &
\\
\cline{1-6}
\multirow{2}{*}{\textbf{\shortstack{Social / \\ Control}}} 
& Bias & 
$1.7$
 & 
$1.6$
 & 
$23.5^{***}$
 & 
$22.3^{***}$
 & 
\multicolumn{2}{c}{\textbf{Social / Control}} & 
\\ 
& Accuracy & 
$25.6^{***}$
 & 
$25.6^{***}$
 & 
$22.6^{***}$
 & 
$22.6^{***}$
 & 
Reported & Prereg &
\\
\cline{1-8}
\multirow{2}{*}{\textbf{\shortstack{Social / \\ Motivated}}} 
& Bias & 
$83.4^{***}$
 & 
$78.5^{***}$
 & 
$31.5^{***}$
 & 
$29.9^{***}$
 & 
$102.4^{***}$
 & 
$96^{***}$
 & 
\multicolumn{2}{c}{\multirow{2}{*}{\textbf{Social / Motivated}}}
\\ 
& Accuracy & 
$17.6^{***}$
 & 
$17.6^{***}$
 & 
$15.1^{***}$
 & 
$15.1^{***}$
 & 
$0.7$
 &
$0.7$
 & 
\\
\cline{1-10}
\\[-\normalbaselineskip]
\multicolumn{2}{r}{\textbf{Mean bias}}
& \multicolumn{2}{c}{0.526
}
& \multicolumn{2}{c}{0.562
}
& \multicolumn{2}{c}{0.512
}
& \multicolumn{2}{c}{0.618
} 
\\
\\[-\normalbaselineskip]
\multicolumn{2}{r}{\textbf{Mean accuracy}}
& \multicolumn{2}{c}{0.604
}
& \multicolumn{2}{c}{0.608
}
& \multicolumn{2}{c}{0.654
}
& \multicolumn{2}{c}{0.646
} 
\\
\end{tabular}
\caption{Experiment 1 results. The values in each 2$\times$2 cell give chi-square statistics for the likelihood-ratio tests between two logistic regressions comparing the relevant conditions. The dependent variable in bias comparisons captures whether a participant chose their marked color (see Materials and Methods, Experiment 1 for a description of marked color), and whether they chose correctly for accuracy comparisons. All models include random intercepts for each replication to control for our dot randomization scheme (see Materials Methods, Experiment 1). Stars indicate statistical significance, with $^{*}p<0.05$, $^{**}p<0.01$, $^{***}p<0.001$ ($p$-values were not adjusted for multiple comparisons). Reported and prereg models are identical, except that prereg models include random intercepts for each participant. Note that the statistics from these two models will be nearly identical when the variance of the participant random intercepts is near zero. The bottom two rows give the proportion of judgments where participants chose their marked color (bias) and the correct color (accuracy) for each condition.}

\label{table:e1_results}
\end{table}

\begin{table}[p]
\centering
\renewcommand{\arraystretch}{1.1}
\setlength\tabcolsep{5pt}
\begin{tabular}{crcccccc}
& & \multicolumn{2}{c}{\textbf{Asocial / Motivated}} \\
& & Reported & Pre-reg 
\\
\cline{1-4}
\multirow{2}{*}{\textbf{\shortstack{Social / \\ Motivated}}} 
& Bias & 
$12.5^{***}$
 & 
$7.6^{**}$
 & 
\multicolumn{2}{c}{\textbf{Social / Motivated}} & 
\\ 
& Accuracy & 
$82.5^{***}$
 &
$26.1^{***}$
 &
Reported & Pre-reg &
\\
\cline{1-6}
\multirow{2}{*}{\textbf{\shortstack{Social / \\ Resampling}}} 
& Bias & 
$0.5$
 & 
$3.8$
 & 
$7.5^{**}$
 & 
$11.3^{***}$
 & 
\multicolumn{2}{c}{\multirow{2}{*}{\textbf{Social / Resampling}}}
\\ 
& Accuracy & 
$59.3^{***}$
 & 
$10.5^{**}$
 & 
$1.9$
 &
$1.9$
 & 
\\
\cline{1-8}
\\[-\normalbaselineskip]
\multicolumn{2}{r}{\textbf{Mean bias}}
& \multicolumn{2}{c}{0.54
}
& \multicolumn{2}{c}{0.567
}
& \multicolumn{2}{c}{0.545
}
\\
\\[-\normalbaselineskip]
\multicolumn{2}{r}{\textbf{Mean accuracy}}
& \multicolumn{2}{c}{0.574
}
& \multicolumn{2}{c}{0.63
}
& \multicolumn{2}{c}{0.621
}
\\
\end{tabular}
\caption{Experiment 2 results. As in Table~\ref{table:e1_results}, the values in each 2$\times$2 cell give chi-square statistics for the likelihood-ratio tests between logistic regression models with and without separate fixed effects for the two conditions. Bias models predicted whether participants chose their motivated color, and accuracy models whether they chose correctly. All models include a fixed effect for motivated color, and a random intercept for each participant. Prereg models include an additional random intercept for each social replication to account for the social yoking scheme (see Materials Methods, Experiment 2). Stars indicate statistical significance, with $^{**}p<0.01$ and $^{***}p<0.001$ ($p$-values were not adjusted for multiple comparisons). The bottom two rows give the proportion of judgments where participants chose their motivated color (bias) and the correct color (accuracy) for each condition. 
}
\label{table:e2_results}
\end{table}

\begin{table}[p]
\centering
\renewcommand{\arraystretch}{1.2}
\setlength\tabcolsep{5pt}
\begin{tabular}{lcc}
& \textbf{Asocial / Motivated} & \textbf{Social / Motivated} \\
\hline
\\[-\normalbaselineskip]
Bias mean & 0.26
 & 0.371
\\
& [0.219
\unskip,
0.303
\unskip] & 
[0.302
\unskip,
0.441
\unskip] \\
\\[-\normalbaselineskip]
Bias standard deviation & 0.075
 &  0.465
\\
& [0.005
\unskip,
0.187
\unskip] & 
[0.359
\unskip,
0.569
\unskip] \\
\\[-\normalbaselineskip]
48 green dots indicator & -0.447
 &  -0.482
\\
& [-0.541
\unskip,
-0.355
\unskip] & 
[-0.61
\unskip,
-0.356
\unskip] \\
\\[-\normalbaselineskip]
49 green dots indicator & -0.229
 &  -0.14
\\
& [-0.327
\unskip,
-0.129
\unskip] & 
[-0.252
\unskip,
-0.029
\unskip] \\
\\[-\normalbaselineskip]
51 green dots indicator & 0.401
 &  0.518
\\
& [0.306
\unskip,
0.499
\unskip] & 
[0.401
\unskip,
0.636
\unskip] \\
\\[-\normalbaselineskip]
52 green dots indicator & 0.703
 &  0.613
\\
& [0.593
\unskip,
0.802
\unskip] & 
[0.489
\unskip,
0.739
\unskip] \\
\\[-\normalbaselineskip]
Green endorsements & & 0.232
 \\
& & [0.198
\unskip,
0.266
\unskip] \\
\\[-\normalbaselineskip]
\hline
\\[-\normalbaselineskip]
Number of samples & 10,000
 & 10,000
 \\
\\[-\normalbaselineskip]
\hline
\end{tabular}
\caption{Parameter estimates from the psychometric models fit to Experiment 1. Each estimate gives the mean sampled value for the parameter averaged over all chains and iterations. Values in brackets below each estimate show cutoffs for the 90\% credible intervals. Individual-level participant biases are not reported.}
\label{table:e1_irt}
\end{table}

\end{refsection}

\end{document}